\documentclass[aip, jap, reprint]{revtex4-1}%
\usepackage{epsfig,pslatex,latexsym,times,amssymb,amsmath,graphicx}
\usepackage{bm, float, lipsum}
\usepackage{amsmath}
\usepackage{amsfonts}
\usepackage{amssymb}
\usepackage{mathrsfs}
\usepackage{color}
\usepackage{graphicx}%
\usepackage{tikz}
\newcommand*\circled[1]{\tikz[baseline=(char.base)]{
            \node[shape=circle,draw,inner sep=1pt] (char) {#1};}}
\providecommand{\U}[1]{\protect\rule{.1in}{.1in}}
\begin{document}
\author{N. J. Harmon}
\email{nicholas-harmon@uiowa.edu} 
\affiliation{Department of Physics and Astronomy and Optical Science and Technology Center, University of Iowa, Iowa City, Iowa
52242, USA}
\author{M. E. Flatt\'e}
\affiliation{Department of Physics and Astronomy and Optical Science and Technology Center, University of Iowa, Iowa City, Iowa
52242, USA}
\affiliation{Department of Applied Physics, Eindhoven University of Technology, P.O. Box 513, 5600 MB, Eindhoven, The Netherlands}
\date{\today}
\title{Organic magnetoresistance from deep traps} 
\begin{abstract}
We predict that singly-occupied carrier traps, produced by electrical stress or irradiation within organic semiconductors, can cause spin blockades and the large room-temperature magnetoresistance known as organic magnetoresistance. The blockade occurs because many singly-occupied traps can only become doubly occupied in a spin-singlet configuration. Magnetic-field effects on spin mixing during transport dramatically modify the effects of this blockade and produce magnetoresistance.We calculate the quantitative effects of these traps on organic magnetoresistance from percolation theory and find a dramatic nonlinear dependence of the saturated magnetoresistance on trap density, leading to values $\sim 20$\%, within the theory's range of validity.
\end{abstract}
\maketitle

\section{Introduction}
The large (up to $\sim 20$\%) room-temperature magnetoresistance in many organic materials (organic magnetoresistance, or OMAR) \cite{Kalinowski2003, Francis2004, Mermer2005a, Prigodin2006, Desai2007, Bloom2007} appears to be a surprising effect of the Pauli exclusion principle on pairs of slowly-hopping carriers.   Quantitative theories of OMAR either involve carriers of the same charge (bipolaron mechanism) \cite{Bobbert2007} or opposite charges (exciton mechanism) \cite{Prigodin2006, Desai2007} occupying the same site, and thus depend nonlinearly on the number of carriers. The dependence of the effect on voltage bias is weaker than would be expected if the effect was driven entirely by carrier-carrier interaction, suggesting that mechanisms that depend linearly on the number of carriers play a role. Furthermore, the effect itself depends sensitively on the age and history of the device,\cite{Niedermeier2008, Baker2012} which makes applications for {\it e.g.} magnetic sensing very challenging. For example, large currents driven over extended time periods  were shown to increase the magnetoresistance from 1\% to over 15\% in poly(para-phenylene vinylene (PPV),\cite{Niedermeier2008} possibly due to charge carrier trap generation caused by extended electrical stress or irradiation. \cite{Niedermeier2008,Bagnich2009, Niedermeier2010, Schmidt2010, Kang2012, Rybicki2012} Optical depletion of trap states in super yellow-PPV decreases OMAR,\cite{Bagnich2009} and X-ray irradiation that produces deep traps with an energy depth $E_t \approx 0.5$ eV in tris(8-hydroxyquinolinato)aluminum (Alq$_3$) strongly enhances OMAR. 
Deactivation of intrinsic traps and introduction of extrinsic traps was also shown to be possible through molecular doping of super yellow-PPV. The net effect was a decrease in the overall magnetic response, which suggests the type of trap (structural or impurity) present is an indicator of the OMAR performance.\cite{Cox2013, Cox2014}
Traps that exhibit strong spin-orbit effects can enhance organic light-emitting diode (OLED) emission and also provide a means for reading out singlet and triplet ratios which could lead to a greater understanding of magnetic field effects in organic semiconductors.\cite{Chaudhuri2013}

\begin{figure}[ptbh]
 \begin{centering}
        \includegraphics[scale = 0.4,trim = 300 250 200 40, angle = -0,clip]{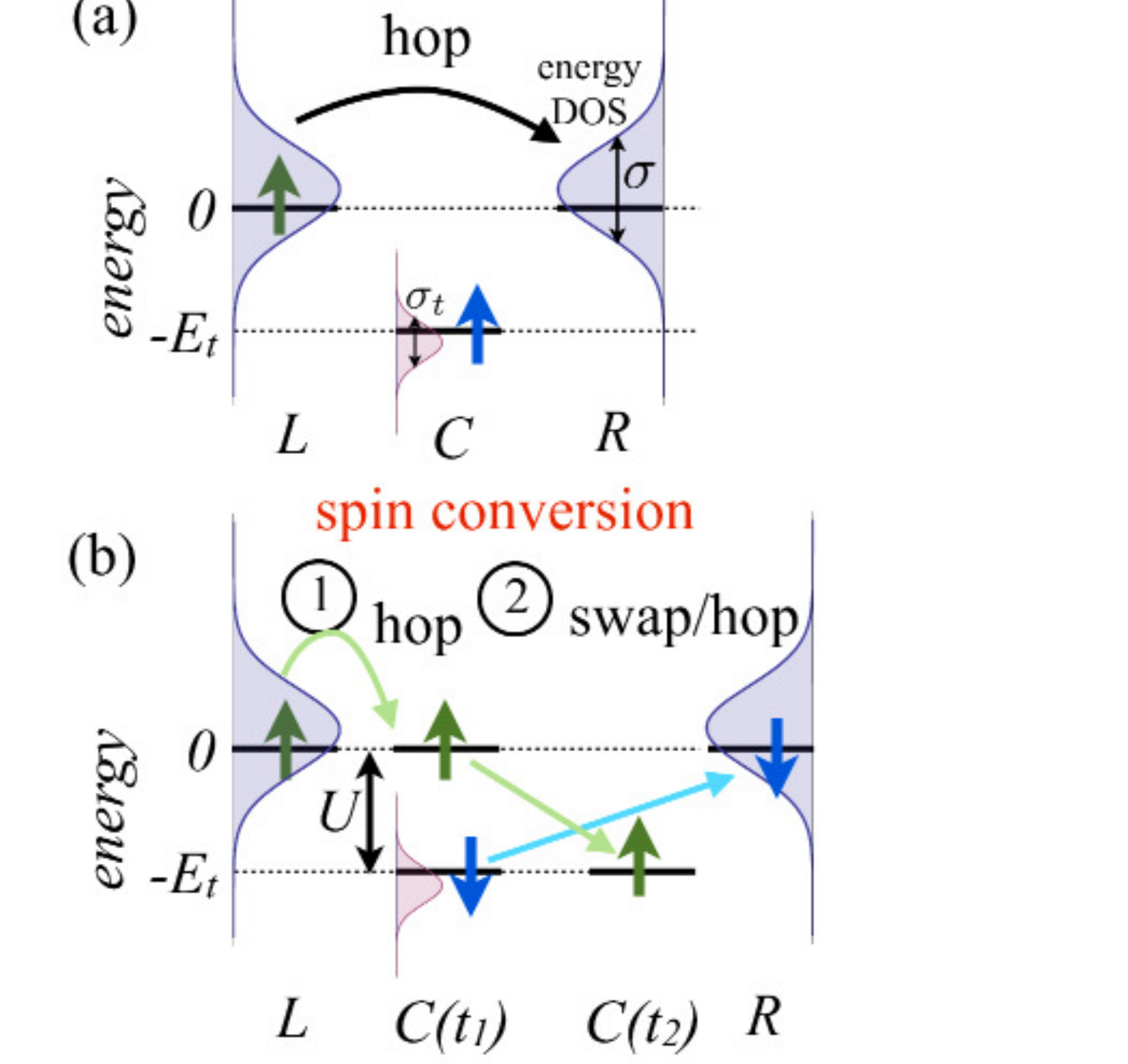}
        \caption[]
{(Color online) Spin blocking in the presence of a trapping site. 
Consider the (green) spin initially positioned at site L to be mobile. 
The (blue) spin located at C is at a trapping site with energy $-E_t$.
Energy of either a vacant/singly occupied or doubly occupied site is shown as a thick line. 
(a) the spin pair does not undergo a spin transition; double occupation at C is forbidden so the spin at L hops to a further site R. 
(b) the spin pair experiences a transition such that the initial triplet configuration becomes a singlet. \circled{1} The spin at L can now hop to C at time $t_1$ though a  Coulomb energy cost of $U$ is incurred.
\circled{2}  At a later time $t_2$, either spin can then hop to site R (in this particular case, the down spin makes the hop). The remaining spin falls to the lower energy state $-E_t$.
Although $U$ is not necessarily resonant with the energy of the trap, there is also energy disorder, with magnitude $\sigma\sim 0.1-0.2$ eV, present which is similar in size to $U$ (Refs. \onlinecite{Helbig1993, Bobbert2007, Wagemans2010}) and $E_t$ which will produce the scenario depicted. Deep traps are highly localizing centers, so the doubly occupied triplet state has an exchange energy assumed to be large enough to neglect the triplet state as a possible intermediate configuration.
}\label{fig:hubbard}
        \end{centering}
\end{figure}

Here we consider  traps in a recent theory \cite{Harmon2012a, Harmon2012b, Harmon2012c, Harmon2012d} of OMAR based on percolation theory \cite{Shklovskii1984}, and surmise that occupied traps are commonly the cause of  OMAR instead of the much more dilute bipolarons or excitons.
Figure \ref{fig:hubbard} provides a qualitative sketch of this trap-induced OMAR theory:
an injected spin-$\frac{1}{2}$ charge carrier or polaron at L encounters a trapped spin at C.
In disordered organic semiconductors, charge transport occurs via tunneling between different localizing sites.
The large exchange energy present for two charges at a single site prohibits the left spin from moving to the center spin's site in a triplet configuration. Instead it must hop to a further, unoccupied site, $R$,  as depicted in Figure \ref{fig:hubbard}(a) \emph{unless} a spin transition occurs.
Spin transitions cause triplet spin configurations to evolve into singlet configurations, which permits the left spin to hop to the center site, as shown in Figure \ref{fig:hubbard}(b).
The Coulomb interaction raises the doubly occupied site at C by an energy $U$. When one of the spins hops off the trapping center at a later time, the other spin solely occupies the trap and returns to the lower energy state.

Several spin-evolution mechanisms permit spin-triplet states to evolve to spin-singlet states, including the different hyperfine fields at the two sites and spin-orbit interactions. These mechanisms are influenced by an applied magnetic field, which thus affects the charge transport and produce magnetoresistance. We assume (1) traps are uniformly, yet randomly, distributed in the organic layer,
(2) current has been driven through the sample for a long enough time that traps are all singly occupied, and 
(3) there is no net spin polarization in either the trapped spins or the hopping spins. We find that trapped polarons create spin blockades in the same manner that it has been believed that free polarons create spin blockades. Traps are a more realistic candidate, however, since free polaron densities are typically very low. Larger trap densities dramatically, and nonlinearly, increase the saturated MR  and lead to values in the range of experimental measurements of OMAR. We note that that nonlinearity yields a three times larger MR at the edge of the regime where the theory is valid. 

\section{Theory}

From the perspective of a semi-mobile hopping polaron, the total concentration of sites and traps is
\begin{equation}
N  = N_0 + N_S + N_T + N^t_0 +  N^t_S + N^t_T,
\end{equation}
where $N_0$ is an unoccupied non-trap site, $N_{S(T)}$ is a site occupied by another polaron forming a singlet (triplet) state, $N^t_0$ is an unoccupied trap, and $N^t_{S(T)}$ is a trap occupied by another polaron forming a singlet (triplet) state.
However not all these sites are possible locations for the polarons to hop to due to the  large exchange energy present for triplet spin pairs and the existence of energetically unfavorable sites.
Instead we define an effective concentration of sites, $N_{eff}'$, that a  spin can hop to:
\begin{equation}
N_{eff}'  = f (N_0 + N_S) + g (N^t_0 +  N^t_S),
\end{equation}
where $f, g \leq 1$ are factors accounting for the reduction of energetically inaccessible sites; $f$ and $g$ are not necessarily equal since they depend upon the amount of energy disorder (a Gaussian distribution is assumed in Fig.~\ref{fig:hubbard}) for normal sites ($\sigma$) and trapping sites ($\sigma_t$) as well as the on-site repulsion energy, $U$, and temperature, $T$.\cite{Osaka1979}
Allowing for spin transitions between singlet and triplet states results in an additional alteration to the effective density of sites:\cite{Harmon2012a, Harmon2012b, Harmon2012c}
\begin{eqnarray}
N_{eff}  &=& f (N_0 + N_S + N_T - N_T (1-\alpha p_{T\rightarrow S} )) \nonumber\\
{}&+& g (N^t_0 +  N^t_S + N^t_T - N^t_T (1-\alpha p_{T\rightarrow S} )).
\end{eqnarray}
The quantity $p_{T\rightarrow S}$, which is the probability for a triplet state to convert to a singlet state, is field, $\omega_0$, dependent and ultimately gives rise to the MR effect. 
It is defined as\cite{Harmon2012c}
\begin{equation}
p_{T\rightarrow S}(\omega_0)  = \frac{1}{3} \Big( 1-  \sum_{m}| P_S^{mm} |^2 \frac{1/\tau_h^2}{(\omega_{m'} - \omega_m)^2 + 1/\tau_h^2}  \Big)
\end{equation}
where $P_S^{mm} = \langle m | P_S | m \rangle$ ($P_S$ is the singlet projection operator), $1/\tau_h$ is the hopping rate, and states $m$ ($\omega_m$)  are the eigenstates (energies) of the total Hamiltonian,
\begin{equation}
\mathscr{H}_{hf} + \mathscr{H}_{Z} 
 = 
 (\bm{\omega}_{hf_1}+ \omega_0 \hat{z}) \cdot \bm{S}_1 +  ( \bm{\omega}_{hf_2} + \omega_0 \hat{z}) \cdot \bm{S}_2,
 \end{equation}
 where $\bm{\omega}_{hf_i} = g\mu_B \bm{B}_{hf_i}/\hbar$ represent the random hyperfine fields present at the two sites.
 $\alpha$ is a factor that accounts for the fact that an encounter with a triplet site temporarily halts the hopping polaron. For instance, even if $p_{T\rightarrow S}  = 1$, there would still be some reduction in the number of accessible sites since $\alpha \leq 1$.
In terms of the actual site density, we write the effective density as
\begin{eqnarray}
N_{eff}  &=& N - (1-f) (N_0 + N_S + N_T) - f N_T (1-\alpha p_{T\rightarrow S} ) \nonumber\\
{}&-& (1-g) (N^t_0 +  N^t_S + N^t_T) - g N^t_T (1-\alpha p_{T\rightarrow S} ).
\end{eqnarray}

By making several assumptions, we can reduce the number of terms in the effective site density.
Here, unlike in previous treatments \cite{Bobbert2007, Harmon2012a}, we consider the density of occupied traps to be much larger than the free polaron density so the possibility of two mobile polarons forming a bipolaron is ignored.
This assumption is justified by the large enhancement of OMAR seen after electrically conditioning the organic material (compared to OMAR in the pristine material) \cite{Niedermeier2008}.
In the steady state, after current has run through the organic sample for a sufficiently long time, most traps are singly occupied and $N^t_0 = 0$. 
This leads to the important conclusion that even if the density of mobile carriers is small, there can still be a significant number of spin blockades due to trapped, immobile charge carriers. 
A final assumption is that the temperature is high enough such that most sites are accessible: $f, g \approx 1$.
This approximation has been observed to be met at room temperature in some organic semiconductors.\cite{Gill1974, Rubel2004} These assumptions allow us to write
\begin{equation}\label{eq:Neff}
N_{eff}  \approx N - N^t_T (1-\alpha p_{T\rightarrow S} ).
\end{equation}
The typical carrier concentration in devices that exhibit OMAR is $10^{20} - 10^{23}$ m$^{-3}$ whereas site concentrations are on the order of $N\sim 10^{27}$ m$^{-3}$.\cite{Shuttle2010}
Estimates of the trap concentration are $10^{23}-10^{24}$ m$^{-3}$ which justifies our neglect of the carrier concentrations in Eq. (\ref{eq:Neff}).\cite{Kumar2003, Cox2013}

The hallmark of a percolation theory of transport in spatially disordered media is the existence of a critical site separation length, $r_c$, paired with a critical resistor $R_c = R_0 e^{2 r_c/\ell}$ with $\ell$ being the localization length.
In the situation at hand, this threshold length is constrained by the bonding criterion: 
\begin{equation}\label{eq:bondingCriterion1}
\int_0^{r_c}4 \pi  N_{eff} r^2 dr = B_c 
\end{equation}
 where $B_c$ tells how many bonds each site must connect to on average to be included in the percolating network; $B_c \approx 2.7$ in three dimensions \cite{Shklovskii1984}.
This simple expression for resistance has been observed in organic semiconductors in the regime of large inter-site separations and high temperatures where the influence of energy disorder is minimized \cite{Gill1974, Rubel2004}.
Typical site separations are 0.5-1.5 nm and localization lengths are $\ell \sim$ 0.1-0.2 nm.

The resistance must be found by first finding $r_c$ by substituting Eq. (\ref{eq:Neff}) into Eq. (\ref{eq:bondingCriterion1}):
\begin{equation}\label{eq:bondingCriterion2}
x \int_0^{y_c}  p_{T\rightarrow S}(\omega_0)y^2 dy - \frac{B_c}{4 \pi \ell^3} + \frac{y_c^3}{3}(N - \frac{x}{\alpha}) = 0,
\end{equation}
with $x = \alpha  N_T^t$ and $y = r/\ell$.
Since the effective density of sites is field-dependent, the resulting critical length $y_c$ is also field-dependent which is the basis for the MR.
For a general hopping rate, $1/\tau_h = v_0 e^{-2 y}$, Eq. (\ref{eq:bondingCriterion2}) can only be solved numerically for $y_c$; the MR is defined as
\begin{equation}\label{eq:mr}
\text{MR} = \frac{e^{2 y_c (\omega_0)}}{e^{2 y_c (0)}} - 1.
\end{equation}
In the limit of slow hopping, which is likely the operative regime in which OMAR is observed,\cite{Baker2012b} $p_{T\rightarrow S}$ becomes independent of the spatial variable:
\begin{equation}
p_{T\rightarrow S}(\omega_0)  = \frac{1}{3} ( 1-  \sum_{m}| P_S^{mm} |^2  )
\end{equation}
so $y_c$ can be solved for as
\begin{equation}\label{eq:slowHopping}
y_c(\omega_0) = \big\langle\left( \frac{3 B_c}{4 \pi \ell^3 (N - x/\alpha + p_{T\rightarrow S}(\omega_0) x)} \right)^{1/3}\big\rangle,
\end{equation}
where angular brackets denote averaging over Gaussian distribution of hyperfine fields (with width $\Delta_{hf}$).

Generalizing to arbitrary hopping rate is straightforward though the calculations are much more time consuming and have been considered elsewhere for bipolarons.\cite{Harmon2012b, Harmon2012c}

Here we use Eq. (\ref{eq:slowHopping}) to calculate the MR instead of making an approximation\cite{Harmon2012c} based on a dilute density of traps.
As a result the MR can be determined as a function of trap density for larger trap densities. 
The more exact numerical calculation here also provides a measure of the accuracy of the approximations leading  to the results of Refs. \onlinecite{Harmon2012b} and \onlinecite{Harmon2012c}:
\begin{equation}\label{eq:fullMR}
\textrm{MR}
\approx 2 \frac{1}{y_{c_1}^2} \frac{x}{N} \int_0^{y_{c_1}}  y^2  \big\langle p_{T\rightarrow S}(0)  - p_{T\rightarrow S}(\omega_0) \big\rangle dy,
\end{equation}
where a renormalized critical length is defined as
$y_{c_1} = y_{c_0} (1-x/\alpha N)^{-1/3}$ and $y_{c_0} = (3B_c/4 \pi \ell^3 N)^{1/3}$ is the critical length in the spinless problem. 
A further approximation is to substitute $y_{c_0}$ for $y_{c_1}$.
The saturated MR can be shown to be approximately
\begin{equation}\label{eq:satMR}
\textrm{MR}_{sat}
\approx \frac{x}{27 N} y_{c_{i}} \qquad \text{for either } i = 0,1.
\end{equation}
\begin{figure}[ptbh]
 \begin{centering}
        \includegraphics[scale = 0.3,trim = 70 170 20 28, angle = -0,clip]{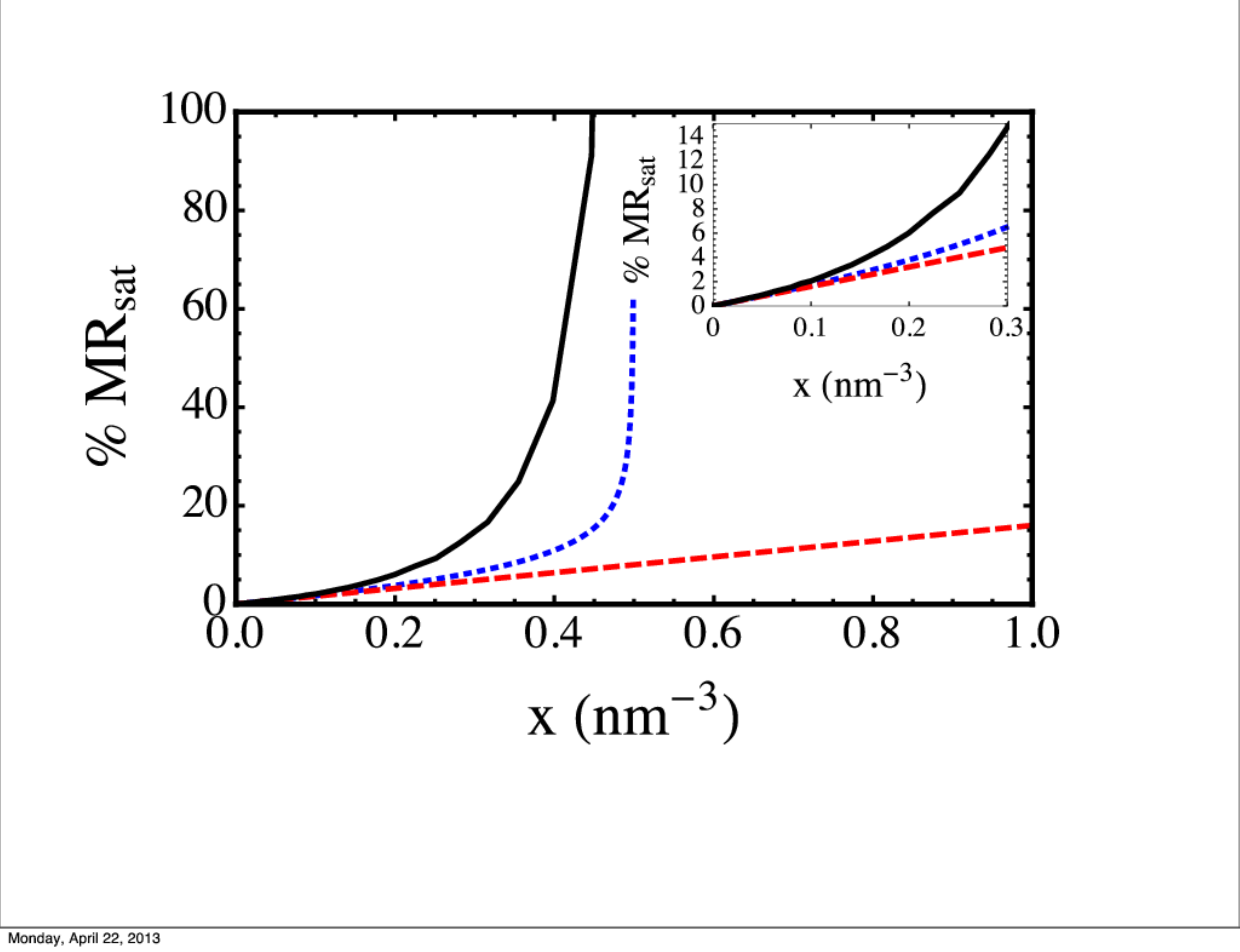}
        \caption[]
{(Color online) Saturated MR percentage versus trap density, $x = \alpha g N_T^t$. Black (solid) line is the calculation using Eq. (\ref{eq:slowHopping}) with Eq. (\ref{eq:mr}).
Blue (dotted) line is from Eq. (\ref{eq:satMR}) using $i = 1$.
Red (dashed) line is from Eq. (\ref{eq:satMR}) using $i = 0$.
Inset shows a smaller horizontal range. 
$N = 1$ nm$^{-3}$, $\ell = 0.2$ nm, $\alpha  = 1/2$, and $g = 1$.
}\label{fig:fig2}
        \end{centering}
\end{figure}

\section{Discussion}
These two different approximations in Eq. (\ref{eq:satMR}), along with numerical calculation using Eq. (\ref{eq:slowHopping}) with Eq. (\ref{eq:mr}), are displayed in Figure \ref{fig:fig2}.
The crudest approximation is to use $i=0$ (red dashed line); the MR trap dependence comes solely from the linear factor seen in Eq. (\ref{eq:satMR}). 
Physically, the MR is increasing linearly with trap density because of the increase in blocking sites. 
The approximation using $i = 1$ behaves differently (blue dotted line); the nonlinear aspect appears since traps also affect MR by effectively decreasing the site density and increasing $y_c$ (cf. Eq. \ref{eq:slowHopping}). Decreasing the site density penalizes dissociative hops severely, which causes a spin blocked site to become a more efficient blockade. This effect  also increases the importance of the hyperfine fields in lifting such blockades.
We find that the more exact numerical calculation (black line) behaves similarly though the MR is larger. The value of the saturated MR grows rapidly upon approaching a critical value, both in the $i=1$ approximation and the numerical calculations; the divergence occurs because the renormalized critical length diverges for $x = \alpha N$ (\emph{e.g.} hopping distance becomes larger since more and more sites are occupied traps). Despite the increase in trap density, the conductivity of the high-conductivity, low-magnetic-field regime remains large.
Unlike the case for other theories \cite{Schellekens2011, Roundy2013, Cox2013}, the nonlinearity in MR with respect to trap density flows naturally from the theoretical approach described here .

It should be remarked that the theory presented herein assumes that blockades are well separated from one another so that more complex situations such as three-spin interactions can be neglected. 
So long as the average trap separation, $r_{sep}$, is larger than the typical hopping length, $r_{av}$, then the trap density is sufficiently low for this approximation to be valid.
A rough estimate can be made; the condition $\frac{4 \pi}{3} x r^3_{sep} \sim 1$ yields $r_{sep} = (\frac{3}{4 \pi x})^{1/3}$ and $\frac{4 \pi}{3} N_{eff} r^3_{av} \sim 1$  yields $r_{av} \approx (\frac{3}{4 \pi N})^{1/3} \times (1 - \frac{x}{\alpha N})^{-1/3}$.
The stipulation that $r_{sep} < r_{av}$ is met when $x \lesssim 0.3$ nm$^{-3} \equiv 3\times10^{26}$ m$^{-3}$. 
In such a case, $x$ is large enough to stray from the low trap density linear regime in Fig.~\ref{fig:fig2}, yielding an MR three times larger than a linear expectation at $x=0.3$ nm$^{-3}$. 
\begin{figure}[ptbh]
 \begin{centering}
        \includegraphics[scale = 0.275,trim = 60 140 140 70, angle = -0,clip]{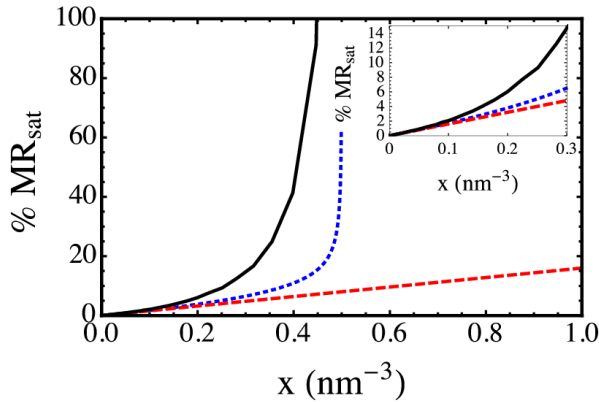}
        \caption[]
{(Color online) MR line shapes for three different trapping densities. $N = 1$ nm$^{-3}$, $\ell = 0.2$ nm,  $\alpha  = 1/2$. $a$ is the width of the Gaussian hyperfine distribution in units of frequency ($a = g \mu_B \Delta_{hf}/\hbar $).
}\label{fig:linshapes}
        \end{centering}
\end{figure}
Figure \ref{fig:linshapes} shows three examples of the MR line shapes with different trap densities. The line shapes are well-fit by Lorentzians and are within the range of experimentally determined OMAR. In both Figures \ref{fig:fig2} and \ref{fig:linshapes} we have assumed a realistic site density for OMAR materials of $N = 1$ nm$^{-3}$.\cite{Shuttle2010}

The theory we have introduced applies to unipolar transport in organic semiconductors. OMAR also occurs in bipolar systems where the constituent pieces are not like-charge pairs but could be electron-hole pairs or excitons. Traps are also important in those systems.\cite{Cox2013} Magnetic field effects similar to OMAR also occur in the photoluminescence and photo-induced absorption but are beyond the scope of our theory.

\section{Conclusion}
In summary, we have shown that neither bipolarons nor excitons are necessary for significant room temperature MR in organic semiconductors; the interaction between a polaron spin and a trapped spin is sufficient to produce efficient transport bottlenecks. The identification of traps as the key sources of OMAR is supported by  recent experiments demonstrating increased MR with electrical stress or x-ray irradiation.

This work was supported by an ARO MURI.

%

\end{document}